\documentclass{aastex61}
\usepackage{mathtools,amssymb,CJK,epigraph,gensymb}
\received{--}
\revised{--}
\accepted{--}
\submitjournal{ApJL}

\shorttitle{\textquoteleft Draconids 2018}
\shortauthors{Egal et al.}

\begin{document}
\begin{CJK*}{UTF8}{gbsn}

\title{The Draconid meteoroid stream 2018: Prospects for satellite impact detection}

\correspondingauthor{Auriane Egal}
\email{aegal@uwo.ca}

\author{Auriane Egal}
\affiliation{Department of Physics and Astronomy, The University of Western Ontario, London, Ontario N6A 3K7, Canada}
\affiliation{Centre for Planetary Science and Exploration, The University of Western Ontario, London, Ontario N6A 5B8, Canada}
\affiliation{IMCCE, Observatoire de Paris, PSL Research University, CNRS, Sorbonne
Universit\'{e}s, UPMC Univ. Paris 06, Univ. Lille}

\author{Paul Wiegert}
\affiliation{Department of Physics and Astronomy, The University of Western Ontario, London, Ontario N6A 3K7, Canada}
\affiliation{Centre for Planetary Science and Exploration, The University of Western Ontario, London, Ontario N6A 5B8, Canada}

\author{Peter G. Brown}
\affiliation{Department of Physics and Astronomy, The University of Western Ontario, London, Ontario N6A 3K7, Canada}
\affiliation{Centre for Planetary Science and Exploration, The University of Western Ontario, London, Ontario N6A 5B8, Canada}

\author{Danielle E. Moser}
\affiliation{Jacobs ESSCA Group, NASA Marshall Space Flight Center, Huntsville, Alabama 35812, USA}

\author{Althea V. Moorhead}
\affiliation{NASA MEO, NASA Marshall Space Flight Center, Huntsville, Alabama 35812, USA}

\author{William J. Cooke}
\affiliation{NASA MEO, NASA Marshall Space Flight Center, Huntsville, Alabama 35812, USA}



\begin{abstract}
Predictions of the 2018 Draconid activity at the Earth and the Sun-Earth L1 and L2 Lagrange points are presented. Numerical simulations of the meteoroids' ejection and evolution from comet 21P/Giacobini-Zinner are performed with a careful implementation of the results analysis and weighting. Model meteoroid fluxes at Earth are derived using as calibration the main peak date, intensity, and shower profiles of previous Draconid outbursts. Good agreement between the model and measurements is found for the 1933, 1946, 1998 and 2011 showers for a meteoroid size distribution index at ejection of about 2.6. A less accurate estimate of the peak time for the 1985, 2005 and 2012 predominantly radio-observed outbursts was found by considering the contribution of individual ejection epochs, while the model peak flux estimate was found to agree with observations to within a factor 3. Despite the promising geometrical configuration in 2018, our simulations predict low Draconid activity is expected on Earth, with a maximum of less than a few tens of meteors per hour around midnight the 9$^\text{th}$ of October, confirming previous models. At the L1 and L2 Lagrange points, however, the flux estimates suggest a ``meteoroid storm''. The Gaia spacecraft at the L2 region might be able to detect small ($\approx \mu$g) Draconid meteoroid impacts centred in a two-hour window around 18$^\text{h}$30$^\text{m}$ UT on the 8$^\text{th}$ of October, 2018. 
\end{abstract}




\section{Introduction}

The October Draconids is an established meteor shower that occurs annually around the 9$^\text{th}$ of October. The shower was linked in 1926 to the Jupiter-family comet 21P/Giacobini-Zinner, known to evolve on a perturbed and ``erratic'' orbit \citep{Marsden1971}. Although the annual activity of the shower is usually low (a few visual meteors per hour), the Draconids produce episodic outbursts (e.g. in 1985, 1998, 2005 and 2011) and meteor storms. In 1933 and 1946, the \textit{Zenithal Hourly Rate} (ZHR) of the shower reached a level of around ten thousand meteors per hour \citep{Jenniskens1995}. In 2012, a completely unexpected Draconid storm was detected by the Canadian Meteor Orbit Radar (CMOR), with an equivalent ZHR reaching 9000 meteors per hour \citep{Ye2014}. The irregularity of these shower displays, caused in large part by the frequent perturbations experienced by the orbit of the comet parent,  make the October Draconids among the most challenging meteor showers to predict.

Past attempts to predict Draconid storms have had mixed success. These typically considered the relative time between the Earth and the comet's nodal passage as well as their nodal distances \citep[e.g.][]{Davies1955}, but no clear correlation between these conditions and the shower intensity was found. For example, in 1972, the close proximity between the Earth and the comet led to predictions of strong activity, but none occurred \citep{Hughes1973}. In contrast, the 2012 radar storm occurred when the Earth crossed the descending node of 21P 234 days after the comet, when no strong storm was expected. The first successful detailed Draconid prediction was the 2011 return, when several observations confirmed the peak times predicted by stream models using numerical simulations \citep[e.g.][]{Vaubaillon2011,Watanabe2008,Maslov2011}. However, the ZHR predictions were more uncertain, ranging from 40-50 \citep{Maslov2011} to 600 \citep{Watanabe2008,Vaubaillon2011}, with some studies even suggesting a possible storm level \citep[ ZHR $\sim$ 7000, cf.][]{Sigismondi2011}. Numerous observations confirmed the timing predictions and revealed an estimated ZHR between 300 \citep{Kero2012} and 400-460 \citep{Trigo2013,Kac2015}.  

In 2018, the Earth will pass within 0.02 au of the descending node of 21P only 23 days after the comet. This encounter geometry seems to favor significant Draconid activity near October 8-9, 2018. However, as several modelers summarized in \cite{Rendtel2017} note, no intense activity has been predicted this year, since the Earth crosses the meteoroid stream in a gap left by previous encounters with the planet. In this study, we present predictions  for the 2018 Draconid return at Earth and in near-Earth space. We use a numerical model of the stream calibrated using the timing and intensity of past Draconid returns to forecast the peak time, peak intensity, and intensity profile of the stream's return on October 8-9.  

\section{Stream Model}  \label{sec:model}

Our Draconid model follows the methodology of \cite{Vaubaillon2005}. The parent comet, 21P, is assumed to be spherical, with a radius of 2 km \citep{Lamy2004} and a nucleus density of  400 $kg.m^{-3}$. The nuclear albedo is taken to be 0.05, with 20\% of the surface being active. The orbital elements of each apparition of the comet are computed (including the influence of the nongravitational forces) from the closest available orbital solution in the JPL Small Body Data Center\footnote{https://ssd.jpl.nasa.gov/sbdb.cgi}.

Meteoroids are ejected during each perihelion passage of 21P between 1852 and 2018 with a time step of one day along the comet orbit for heliocentric distances less than 3.7 AU \citep{Pittichova2008}. A total of 460 000 particles are ejected at each apparition of the comet, covering the size bins $[10^{-4},10^{-3}]$ m : (10$^{-9}$,10$^{-6}$) kg (160 000 particles), $[10^{-3},10^{-2}]$ m : (10$^{-6}$,10$^{-3}$) kg (170 000 particles) and $[10^{-2},10^{-1}]$ m : (10$^{-3}$,1) kg (130 000 particles). In total, the simulated stream consists of 12 million test particles. The meteoroid density is taken to be 300 kg.m$^{-3}$ based on \cite{Borovicka2007}.

The particles are ejected isotropically from the sunlit hemisphere of the comet. The ejection velocities follow the \cite{Crifo1997} model, which produces particles with low ejection velocities in better agreement with recent Rosetta measurements \citep[e.g.][]{Fulle2016}. Each simulated meteoroid is integrated in time using a $15^{th}$ order RADAU integrator \citep{Everhart1985}. The gravitational attraction of the Sun, the eight planets and the Moon, as well as general relativistic corrections are taken into account. Solar radiation pressure and Poynting-Robertson drag are also included. The Yarkovsky-Radzievskii effect was ignored as it is negligible for particle sizes $<$10 cm \citep{Vokrouhlicky2000}.

Meteoroids are considered to be potential impactors if they pass within $\Delta X=V_r \Delta T$ of the Earth, where $V_r$ is the relative velocity between the planet and the particle. The time criterion $\Delta T$ depends of the shower and  is taken here to be one day (for the Earth and a Draconid meteoroid, $\Delta X \simeq 1.15\times10^{-2}$ AU). To estimate the shower flux, the number of  simulated particles is scaled to reflect the number of meteoroids that would be released by the comet. Thus each simulated particle is assigned a weight $W$ which is the number of ``real'' meteoroids it represents. This weight depends on several parameters, such as the cometary activity or the size, direction and speed of the particles ejected. Here the weights follow \cite{Vaubaillon2005}, with various improvements. These include: the cometary gas and dust production rates evolve with the comet's heliocentric distance, and follow from the results of telescopic observations of 21P performed by the NASA Meteoroid Environment Office in 2011 \citep{Blaauw2014} and 2018 (in progress). As well, the ice sublimation to dust production rates are not assumed to be constant with time. These improvements allow incorporation of variable comet activity with  heliocentric distance, and disentangle the comet's dust and gas production. Detailed equations including the final weighting expression will be presented in a forthcoming publication.
  
  \subsection{Modelling the shower flux}
  
  The fraction of simulated meteoroids striking the Earth is very small. To derive a reliable estimate of the meteoroid flux and shower duration, we associate the number of meteoroids contributing to the shower as the number inside a sphere $\mathcal{S}$ centered on the Earth. This sphere has a radius $R_s$ chosen to be large enough to give good statistics but small enough ($<<\Delta X$) to yield a reasonable shower intensity and duration. In this work, $R_s$ is fixed to $V_\oplus\delta t$, with $V_{\oplus}$ the Earth's velocity and $\delta t$ a time parameter of 1 hour, of order of the duration of the core activity for Draconid storms ($R_s\leq 20$ Earth radii). If this distance criteria does not contain a total number of simulated impactors $\ge$10, $\delta t$ is increased until, in the extreme case, it reaches 6h (i.e. slightly longer than the full duration of a typical Draconid shower). The meteoroid flux density $\mathcal{F}$ is obtained by multiplying the spatial density in $\mathcal{S}$ with the relative velocity between the Earth and the meteoroids \citep{Arlt1999,Brown2000,Vaubaillon2005}. 
  
  The \textit{Zenithal Hourly Rate} (ZHR), a measure of the expected number of meteors a visual observer could see under near-ideal conditions, is related to the flux density using \cite{Koschack1990}:
  \begin{equation} \label{eq:ZHR}
   ZHR=\frac{37\mbox{ }200\mbox{ }\mathcal{F}}{(13.1r-16.5)(r-1.3)^{0.748}}
  \end{equation}
 where $r$ is the population index of the observed shower. Our simulated activity profiles are derived from the flux estimate using Equation \ref{eq:ZHR} and the model peak time is estimated by computing a weighted average of the ZHR evolution. 

\section{Validation and flux calibration} \label{sec:calibration} 

Our approach depends on certain parameters that have not been specifically measured for 21P/Giacobini-Zinner. Among these, the most important is the size distribution index $u$ of the meteoroids at ejection, which relates the cumulative number of meteoroids $N$ with sizes $\ge$ a as $N=Ka^{-u}$, $K$ being a normalization constant. The choice of $u$ impacts the shape, length, and intensity of the activity profiles predicted by the simulations. It may be estimated from previous Draconid observations or from \emph{in situ} measurements. To estimate $u$, we compared observed and simulated ZHR profiles obtained with our method, with the goal of validating and calibrating our flux determination for the shower. 
 
\subsection{ZHR calibration}
  
\begin{figure}
 \centering
  \includegraphics[width=0.86\textwidth]{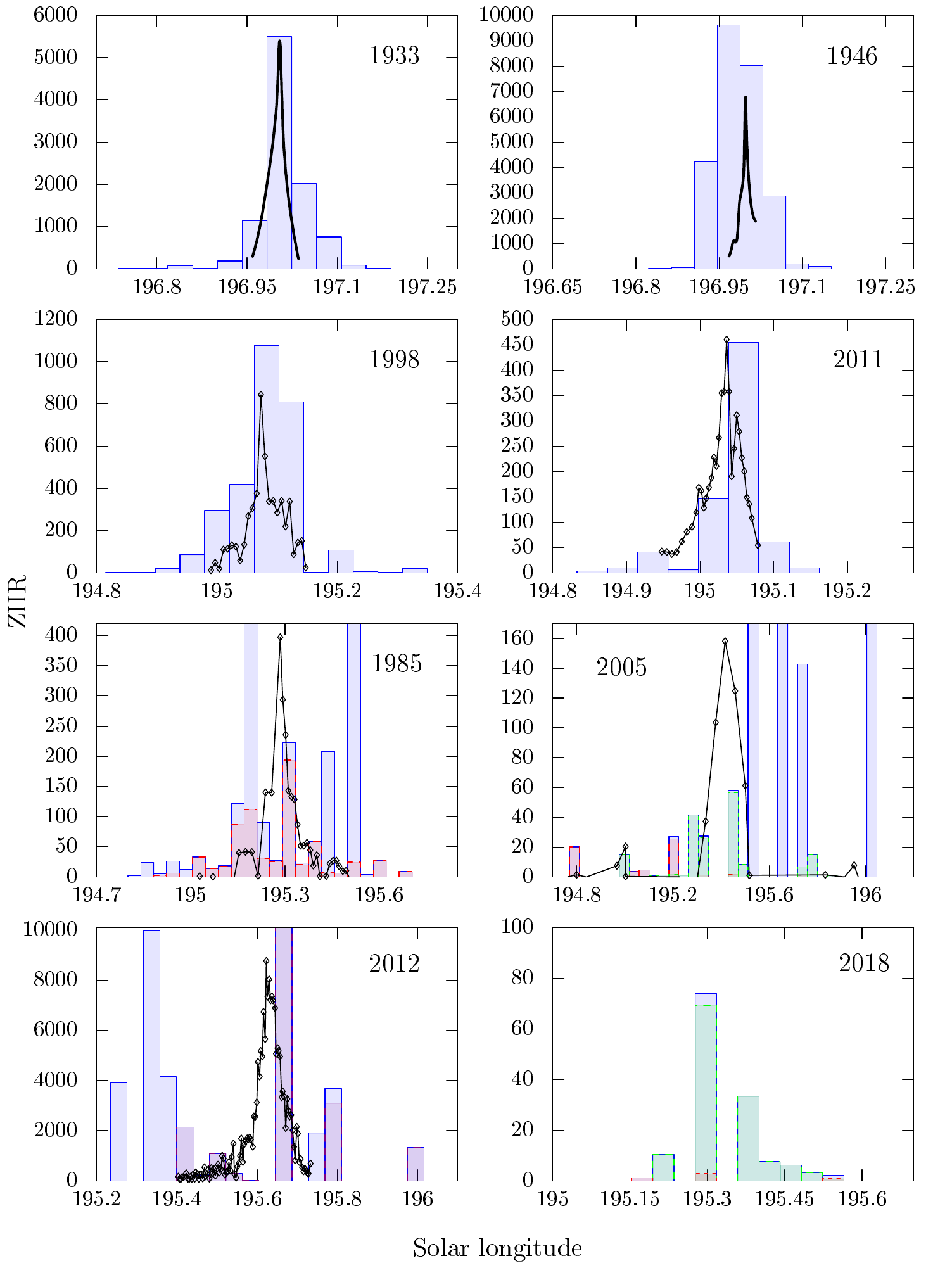}\\
   \includegraphics[width=0.66\textwidth]{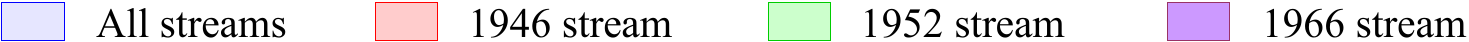}
   \caption{\label{fig:calib_final} Observed (black solid curve) and simulated (colored boxes) ZHR profiles of 7 Draconid outbursts. References for the observations are respectively (from upper left chronologically) \cite{Watson1934,Kresak1975,Watanabe1999,Kac2015} and \cite{Simek1986,Campbell-Brown2006,Ye2014}. The right-bottom plot represents the predicted 2018 activity curve derived with the method presented at section \ref{sec:model}.}
 \end{figure}

 Since our approach aims to reproduce not only the time of the shower maximum but also its duration and intensity, we need to calibrate the simulated ZHR profiles obtained with Equation \ref{eq:ZHR} against Draconid observations. The best agreement between our simulations (blue boxes) and the observations (black curves), presented in Figure \ref{fig:calib_final}, was found for a size distribution index $u$ of 2.64 at ejection. Though this value is low, such an estimate of $u$ is not surprising for a shower as irregular as the Draconids. Since we lack any direct or indirect measurement of $u$ with which to compare our value, we fix it to 2.64 for the rest of the study. 
 
 The ZHR is derived from the meteoroid flux assuming a population index $r$ of 2.6, determined for the 2011 \citep{Toth2012,Kac2015} and 1933 \citep{Plavec1957} showers. In the calibration process, $r$ is assumed to be constant in order to determine the weighting solution that best reproduces all the outbursts, which reinforces the reliability of the shower's predictions. 
 
 \subsubsection{Validation}
 
Our simulations show enhanced Draconid activity in the visual meteoroid size range in 1933, 1946, 1998 and 2011, and for smaller meteoroid detectable as radio outbursts in 1985, 2005 and 2012 \citep{McIntosh1972,Hughes1973,Campbell-Brown2006}. However, we also predict a non-observed moderate activity in 1940 and 1953, and miss the radio enhancement detected by the Jodrell Bank radar in 1952. Our results are similar to those presented in \cite{Kastinen2017}. 
 
 For the 1933, 1946, 1998 and 2011 Draconid returns (cf. Figure \ref{fig:calib_final}), the large number of particles impacting Earth allow the peak time, intensity and duration of the shower to be determined without difficulty. For the 1985, 2005 and 2012 outbursts, the small number of particles impacting the sphere $R_s$ forced us to increase the sphere radius $R_s$ to $\delta t=6h$. In this situation, only the trails whose nodes intersect the Earth's orbit were considered (i.e. the 1946 trail in 1985, the 1946 and 1952 trails in 2005 and 2018, and the 1966 trail in 2012).
 
 The shower parameters, as derived from the observations and the simulations, are summarized in Table \ref{table:comp_obs}. The modelled peak times agree with measurements to within less than half an hour in most cases, and to within better than 10 minutes for the outbursts dominated by visual-sized meteoroids. The intensity and duration estimates are also in good agreement with the observations. The low number of particles available to compute the ZHR of the other Draconid returns does not permit reliable determination of the peak time and intensity of those showers. Figure \ref{fig:calib_final} illustrates the nonphysically long duration that would be predicted for these shower returns if we consider all the incoming streams. By considering the influence of a few specific comet ejection epochs instead, we were able to determine the peak maximum with a difference of about half an hour and an intensity estimate correct within a factor of 2 or 3. However, the need to  appeal to a different methodology to reproduce these returns highlights the lower reliability of our predictions for Draconid showers caused principally by small particles detectable by radar. 

\begin{table}[!ht]
\centering
 \begin{tabular}{|c|c|c|c|c|c|c|c|}
    \multicolumn{1}{c}{ } & \multicolumn{3}{c}{Optical observations} & \multicolumn{4}{c}{Simulations}\\[0.1cm]
   \hline
 Date & Time (UT) & Duration & ZHR & Time & Duration & ZHR & Time*\\
 \hline
 \mbox{ }9/10/1933$^{1}$ & 20h15 & 4h30 & 5400 to 30 000
           & 20h23 & 4h30 & 5500 & 20h08\\
 10/10/1946$^2$ & 3h40-50 & 3-4h & 2000 to 10 000 & 3h34 & 4h & 9650 & 3h38\\
 \mbox{ }8/10/1998$^3$ & 13h10 & 4h & 700 to 1000 & 13h17 & 4h & 1075 & 13h20 \\
 \mbox{ }8/10/2011$^4$ & 20h00-15 & 3-4h & 300-400 to 560 & 20h17 & 4h30 & 450 &  20h05\\
 \hline 
  \multicolumn{8}{c}{}\\[-0.2cm]
   \multicolumn{1}{c}{ } &  \multicolumn{3}{c}{Radio observations} & \multicolumn{4}{c}{Simulations}\\[0.1cm]
 \hline
  Date & Time (UT) & Duration & ZHR & Time & Stream & ZHR & Time*\\
  \hline
 \mbox{ }8/10/1985$^5$ & 9h25-50 & 4h30 & 400 to 2200 & 9h54 & 1946 & 180 & 10h30 \\
 \mbox{ }8/10/2005$^6$ & 16h05 & $\ge$ 3h & 150 & 14h32 & 1946 & 20& 17h20 \\
               &       &          &     & 16h02 & 1952 & 60 & - \\
 \mbox{ }8/10/2012$^7$ & 16h40 & 2h & 9000 & 17h15 & 1966 & 20 000 & 15h56 \\
 \hline             
 \end{tabular}
 \caption{\label{table:comp_obs}Comparison of observations with model predictions for Draconid outbursts/storms. Observation sources: $^1$[\cite{Watson1934,Olivier1946,Cook1973}], $^2$[\cite{Lovell1947,Kresak1975,Hutcherson1946,Jenniskens1995}], $^3$[\cite{Koseki1998,Arlt1998,Watanabe1999}], $^4$[\cite{Toth2012,Kero2012,Koten2014,Molau2014,Kac2015,Trigo2013}], $^5$[\cite{Simek1986,Simek1994,Lindblad1987,Mason1986}],$^6$[\cite{Campbell-Brown2006,Koten2007}],$^7$[\cite{Ye2014}], *Peak time estimated from the stream median location, cf. \ref{sec:2018Earth}.}
\end{table}

\section{2018 Draconids at Earth}

\subsection{Activity profile} \label{sec:2018Earth}

\begin{figure}[!ht]
   \centering
   \includegraphics[width=\textwidth]{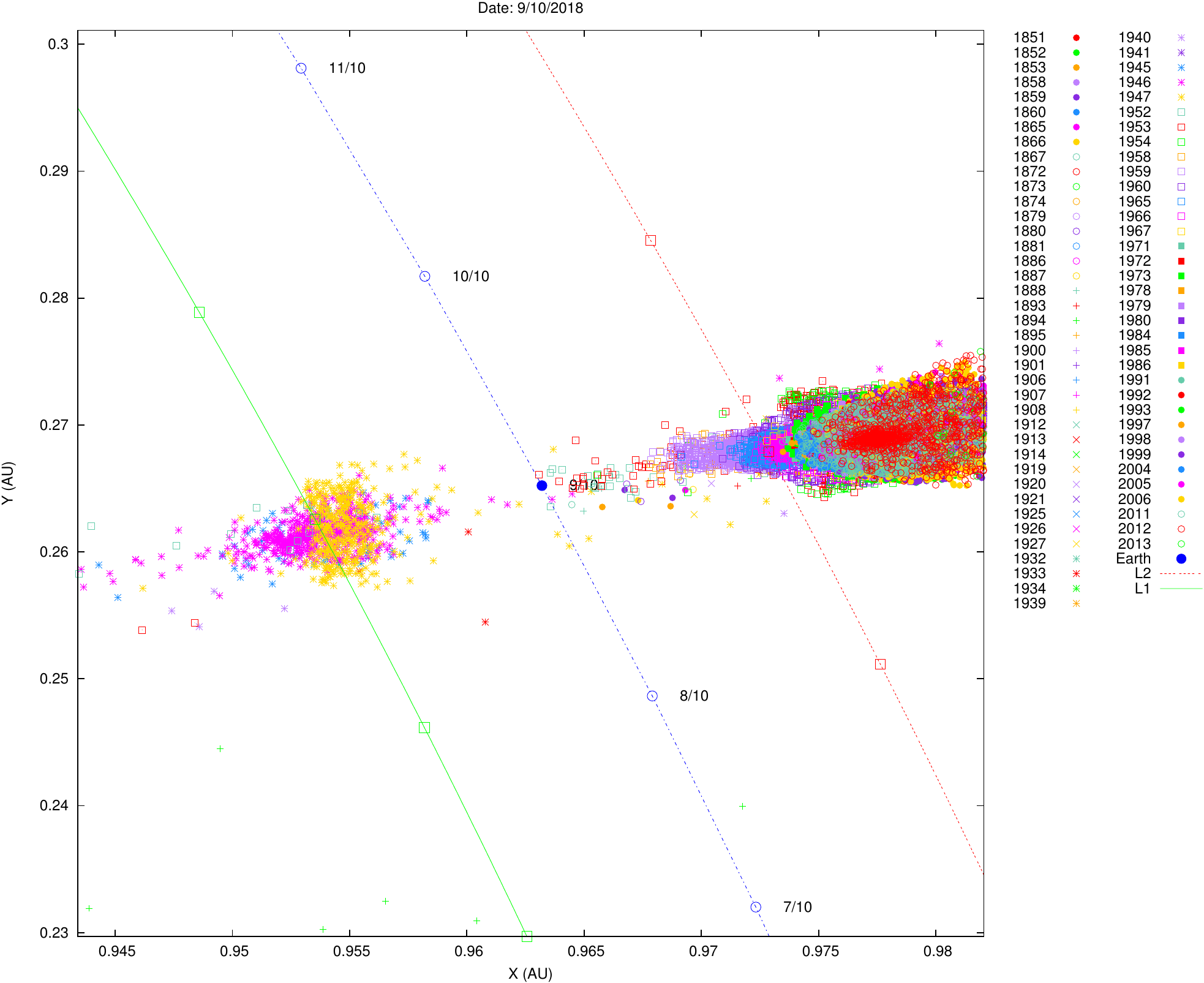}
   \caption{\label{fig:Nodes_2018} Draconid meteoroid nodal crossings close to the Earth's orbital plane on 8-9 of October 2018. Each symbol indicates a particles ejection epoch (legend) while the Earth's path is shown in blue with L1 in green and L2 in red.}
  \end{figure}

Though the geometry between 21P and the Earth looks promising in 2018, our simulations confirm that the planet will cross the meteoroid streams through a gap left between the 1946 and 1952 trails (cf. Figure \ref{fig:Nodes_2018}). Because of the small number of particles which are usable for the flux computation, we  again adopt the strategy applied to the radio showers. The predicted activity profile is presented in Figure \ref{fig:calib_final}, and is mainly composed of particles released during the 1952 perihelion passage of 21P and which would be observable in the visual range. A minor contribution from the 1946 stream consists mainly of smaller particles that would be detected by radar. From this histogram, we estimate the maximum of the shower will occur around 22h20 the 8$^\text{th}$ of October ($L_\odot\sim195.327\degree$ with a half an hour uncertainty) and with a ZHR not exceeding a few tens of meteors per hour.

 However, these small-numbers statistics ($<$10 particles) motivates us to determine the potential peak time with another method. The last column of Table \ref{table:comp_obs} presents the peak times associated to the closest approach date between the Earth and the median position of the meteoroid streams \citep{Vaubaillon2005}. With this approach, we reach an accuracy of less than 10 minutes for the visual outbursts, and about one hour for the other (radar) returns. Applying this technique to 2018 leads to an estimated maximum around 23h51 ($L_\odot=195.390\degree$) on October 8$^\text{th}$, in good agreement with other predictions \citep{Rendtel2017}.
 
 \subsection{Predictions from other models}
 
Several modelers have made predictions for the 2018 Draconids; a summary of these is shown in Table \ref{table:other_models} alongside the results for this work.  Peak timing predictions range from 22h20 on October 8$^\text{th}$ to 00h30 on October 9$^\text{th}$. ZHR predictions are varied, with the majority of models indicating ZHRs on the order of 10-50 meteors per hour. Two exceptions, \cite{Ye2014} and \cite{Kastinen2017}, predict outbursts in 2018.  But numerical simulations using the MEO's MSFC Meteoroid Stream Model \citep{Moser2008}, similar in very broad strokes to the model presented here, indicate that the Earth will pass through a gap in the stream resulting in little activity at the Earth but significant flux near L2 \citep{Moser2017}.

\begin{table}[!ht]
\centering
 \begin{tabular}{|c|c|c|c|c|}
 \hline
 Modeler & Trail & $L_\odot$ & ZHR & Comment \\
 \hline
 Egal$^{1}$ & Mult. & 195.327 & 10s & This work, see text \\
 Maslov$^{2}$ & 1953 & 195.354-195.395 & 10-20 & Rarified, no strong outburst expected \\
 Vaubaillon$^{3}$ & - & 195.374 & 15 & - \\
 Maslov$^{3}$ & 1953 & 195.378 & 10-15 & - \\
 Egal$^{1}$ & Mult. & 195.390 & 10s & This work, alternative peak method \\
 Ye$^{4}$ & - & 195.4 & - & Nodal footprint offset, but outburst similar to 2012 possible \\
 Kastinen $\&$ Kero$^{5}$ & - & 195.4 & - & Could be up to twice as large as 2011/2012 outbursts \\
 Sato$^{3}$ & 1953 & 195.406 & 20-50 & Dust spread out \\
 Vaubaillon$^{6}$ & Mult. & 195.415 & 15 & - \\
 NASA MEO$^{7}$ & Mult. & 195.416 & - & Activity expected to be mild to moderate \\
  \hline             
 \end{tabular}
 \caption{\label{table:other_models}2018 Draconid model predictions at Earth from various modelers.  Sources: $^1$[This work.], $^2$[\cite{Maslov2011}], $^3$[\cite{Rendtel2017}], $^4$[\cite{Ye2014}], $^5$[\cite{Kastinen2017}],$^6$[Vaubaillon, personal comm.],$^7$[\cite{Moser2017}]}
\end{table}

\section{2018 Flux at L1 and Gaia}

From Figure \ref{fig:Nodes_2018}, we see that while low Draconid activity is expected for the Earth in 2018, this is not the case for the L1 and L2 Lagrange points. The Draconid meteoroid flux at these positions is sufficient to warrant clos99ier examination of the risks incurred by spacecrafts located there. In this section, we follow exactly the same methodology presented in \S\ref{sec:model}, with the calibrated weights determined in \S\ref{sec:calibration} to estimate the meteoroid fluxes. Although ZHR values imply an Earth-bound visual observer and therefore do not apply to L1 and L2, for ease of comparison we provide an equivalent ZHR estimate to contrast with previous outbursts on Earth.

In Section \ref{sec:calibration}, we calibrated our simulated ZHR profiles using four observed Draconid outbursts using Equation \ref{eq:ZHR}, which corrects the flux estimate to include only meteoroids that would produce a meteor brighter than magnitude +6.5 \citep{Koschack1990}. We then constrained our flux estimate to particles of a mass higher than 10 mg, and a radius above 2 mm for our selected density. All the flux estimates presented in this work use these particle mass limit. 

\begin{figure}[!ht]
 \centering
 \includegraphics[width=.9\textwidth]{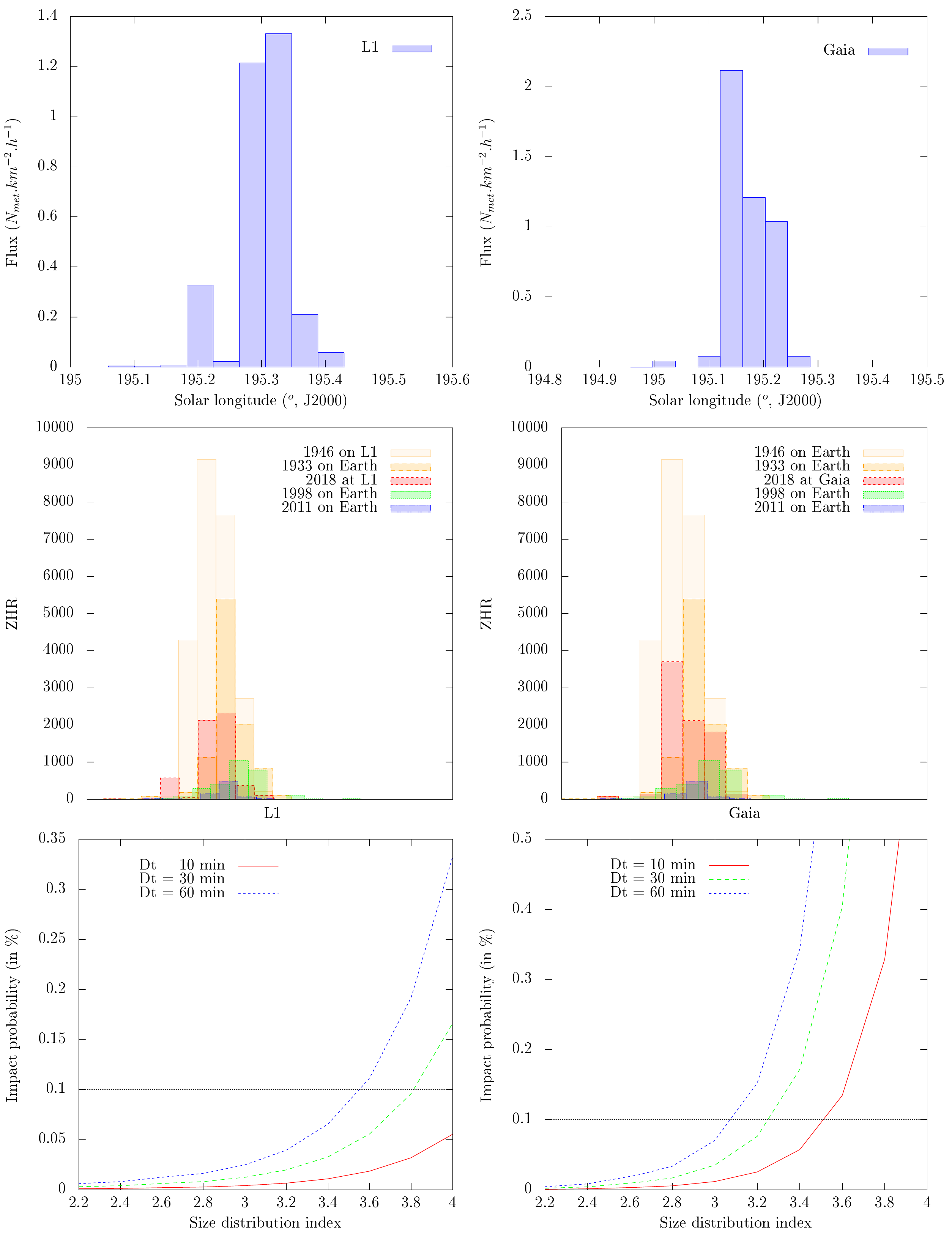}\\
 \caption{\label{fig:flux_L1_gaia} Modelled meteoroid flux (top panel) and equivalent ZHR (middle panel) estimated around the L1 region (left) and the Gaia spacecraft (right) at L2. The bottom panel illustrates the impact probability between a meteoroid of mass $\ge$ 10 mg and a spacecraft of 100 m$^2$ for different size distribution indexes and shower durations.}
\end{figure}

\subsection{L1}

The L1 region hosts a significant number of active spacecraft (e.g. SOHO, ACE, GGS and DSCOVR) that might be impacted by meteoroids released by the comet in 1946. Figure \ref{fig:flux_L1_gaia} presents the model flux profile around L1, and an equivalent ZHR compared with previous Draconid showers on Earth. The high number of particles selected for the flux computation ($\ge$50) provides reliable profiles. We estimate that the L1 surroundings will see enhanced Draconid activity during the period $L_\odot=[195.264\degree-195.368\degree]$, with lower activity between $L_\odot=195.224\degree$ and $L_\odot=195.409\degree$. The peak maximum should occur at $L_\odot=195.308\degree$, reaching a maximum flux of 1.33 $\text{meteoroid}.\text{km}^{-2}.\text{h}^{-1}$. The equivalent ZHR of about 2350 is more than twice that of the 1998 Draconid outburst on Earth.

   \subsection{Gaia} 

Because the Gaia spacecraft is the only mission currently in operation orbiting around the L2 point, we chose to determine the meteoroid flux directly at the spacecraft. The Gaia ephemeris was provided by JPL's HORIZONS system.  Our simulations produce more than 190 impactors useable for flux computation, hence the activity at Gaia is expected to be higher than at L1 (cf. Figure \ref{fig:flux_L1_gaia}). The spacecraft will mostly encounter particles ejected in 1959, with minor contributions from the 1965 and 1972 trails. The main activity will occur in the interval $L_\odot=[195.112\degree-195.224\degree]$, with activity continuing between 195.080$\degree$ and 195.286$\degree$. The peak maximum, estimated at $L_\odot=195.172\degree$, is reached for a flux of more than 2 $\text{meteoroids}.\text{km}^{-2}.\text{h}^{-1}$. This intensity ($\text{ZHR}_\text{equivalent}\sim3700$) is about four times higher than the 1998 outburst on Earth, and is comparable to the historic 1933 peak ZHR estimated by \cite{Watson1934}. Thus, we predict a level of activity at both L1 and L2 equivalent to a meteor storm.

\section{Discussion}

The Draconids are complicated to predict. Multiple close encounters between the parent comet and Jupiter, in addition to sudden modifications of its nongravitational forces (NGF), prevent an accurate ephemeris for 21P prior to 1966. Around the 1959 apparition of the comet, the NGF coefficients changed significantly, perhaps because of the activation of discrete source regions at the comet's surface \citep{Sekanina1993}. While the comet ephemeris in 1959 does not appear to influence our results, the comet could have been particularly active at this return and our flux determination might underestimate the activity expected at Gaia in 2018. 

The uncertainty of the shower's intensity as derived from meteor observations lead to a factor of 2-3 uncertainty in our activity predictions. Despite the good agreement between our simulated activity profiles and the visual showers, we were not able to fully reproduce all the radio outbursts in terms of peak time, intensity, and duration to within the same uncertainty range. This dichotomy raises the question of the validity of the meteoroid size distribution considered at the ejection ; discrepancies between the activity recorded from radar and optical observations support the idea that the particles size distribution at the comet may not be well reproduced by a single power law function.

The predicted Draconid activity at L1 and Gaia is highly dependent on the ejecta size distribution index $u$, which was never measured for comet 21P/Giacobini-Zinner and is truly uncertain. Therefore, Figure \ref{fig:flux_L1_gaia} presents the impact probability (in \%) for Draconid meteoroids of mass $\ge$10 mg and a spacecraft of 100 m$^2$ ($\sim$ GAIA's sunshield surface) for different $u$ values and shower peak durations. Even if we are not inclined to consider large $u$ parameters ($>$3.5), this plot illustrates how the Draconid threat in these regions can increase with a slight modification of the size distribution index. Though the mass index is uncertain at very small sizes, simple extrapolation of our model result suggests that Gaia could expect several impacts from Draconids with mass of order $\approx$ 1 $\mu$g during the course of the storm. As a result, an impact analysis at Gaia could produce the first \emph{in situ} measurement of the smallest Draconid meteoroids.

\section{Conclusion}

This work presents our predictions for the 2018 Draconid flux at Earth and in near-Earth space. A numerical model of the meteoroid stream was implemented by updating and adapting the methodology of \cite{Vaubaillon2005} to comet 21P/Giacobini-Zinner. The simulated meteor showers were successfully calibrated using the peak time, intensity, and activity profile of four visual Draconid outbursts. With the same parameters, a fair estimate of the date of the other showers, caused by radar-sized particles, was provided. Predictions for 2018 suggest a maximum activity on Earth of a few tens of meteors per hour, around 00h-00h30 the 9$^\text{th}$ of October. Because of our model's limitations, the activity caused by small particles is still uncertain, even if we are not expecting a storm as occurred in 2012. However, satellites located at the L1 and especially the L2 region will probably experience intense meteoroid activity, with fluxes reaching more than 2 km$^{-2}$.h$^{-1}$ for particles with masses higher than 10 mg, and an equivalent ZHR of at least 3700 meteors per hour.

\acknowledgments{We thank J. Vaubaillon for his support and advice regarding his model's implementation. This work was carried out under NASA Meteoroid Environment Office Cooperative agreement 80NSSC18M0046.}

%

\vspace{5mm}

\end{CJK*}



\end{document}